\documentclass[useAMS,usenatbib]{mn2e}

\usepackage[dvips]{graphicx}

\title[Polytropic Isothermal Gas Spheres]{A Polytropic Approach to Semi-relativistic Isothermal Gas Spheres at Arbitrary Temperature}
\author[de Sousa {\rm \&} de Araujo]
  {Claudio M. G. de Sousa, $^{1,2}$\thanks{claudio@unb.br , claudiogomes@ufpa.br}
  Evandro A. de Araujo. $^{1}$\thanks{evandro.ufpa.fisica@gmail.com}
  \\
  $^1$ Faculdade de Fisica Ambiental, Universidade Federal do Oeste do Para, 
       Av. M. Rondon, s/n, Santarem, PA 68040-070, Brazil.
\\
   $^2$  Diretoria de Fisica, Universidade Catolica de Brasilia,
         QS 07 Lt 01 EPCT Aguas Claras,
         Brasilia, DF 71966-900, Brazil.
  }
  
\date{Released 2011 Xxxxx XX}

\pagerange{\pageref{firstpage}--\pageref{lastpage}} \pubyear{2011}

\def\LaTeX{L\kern-.36em\raise.3ex\hbox{a}\kern-.15em
    T\kern-.1667em\lower.7ex\hbox{E}\kern-.125emX}

\begin{document}

\label{firstpage}

\maketitle

\begin{abstract}
We use standard polynomial expansion technique to show the existence of a relation between polytropic model and 
the description of gas spheres at finite temperature. A numerical analysis is made concerning the obtained perturbative parameters.
It is shown that there is a correspondence between polytropic and gas sphere thermal models for fermions and bosons. For instance, the polytropic index $n$ displays evident correlation with temperature and chemical potential. 

\end{abstract}



\section{Introduction} \label{s:Intro}

As they barely emit light, compact objects are considered as part of the dark matter in cosmology.
Dark matter is a parcel of mass that can only be detected by its gravitational effects. With this definition and depending on where the referential is placed many things can be considered as dark matter; for instance, even Jupiter can be considered as dark matter for a distant observer. 
This subject has increasingly been accepted as decisive for understanding the evolution of the universe.
There are even speculations \cite{Portilho} that in the Earth, the 
Chandler wobble excitation and damping, one of the open problems in geophysics, can
be considered as a consequence of geological interaction with an oblate ellipsoid made of dark matter.
The same phenomenon is also observed in other planets.

Differently from dark energy, which has no effective mass but presents its effects similarly decisive 
for the evolution of the universe,
dark matter has mass and may consist of many kinds of particles: dust, neutrinos, neutrons, protons (hydrogen), and even bosons, like alpha particles, Higgs bosons or axions \cite{Kolb}.
Many of those particles come from the early universe, encompassing some of the lore of its evolution.
Fraction of those particles are roving over the free space. But, in the early universe, some of those particles departed from roaming, clustered and formed self-gravitating compact objects, like fermion stars and boson stars.

Fermion stars is a general name denoting particular ones like neutron star and white dwarf stars.
Since Oppenheimer and Volkoff \cite{Oppenheimer}, these compact objects 
have received large attention and many of their properties are already determined, and
astronomers can use the theoretical information to detect them.
From Chandrasekhar \cite{Chandrasekhar} one can also learn that neutron stars
and white dwarf stars can be studied as a degenerate Fermi gas under
itself gravitational field, with an equation of state to determine its
pressure and density of energy.

In contrast to fermion star there is the so-called boson star \cite{Ruffini, Liddle, Mielke2}, 
built up with self-gravitating bosons at zero temperature. 
In this case, one can also use
relativistic approach, but the star structure can be studied directly from
the stress-energy tensor since it is possible to write a Lagrangian for
bosons in place of using an equation of state.
Despite this compact object has not been detected yet, there are many theoretical
efforts to understand its properties, like formation and stability \cite{Gleiser, Mielke1},
rotation \cite{Claudio95, Claudio06}, and even the interaction \cite{Claudio98}
between bosons and fermions in the same spherical system.

The aim of the present article is to show that there is a relation between the statistical mechanics of the gas inside the star and the polytropic model for both fermions and bosons, starting from the isothermal gas sphere, and departing from this stage to slightly reach a non-zero temperature in first approximation.


\section{Systems of self-gravitating bosons and fermions}\label{s:Systems}
It is well known that neutron stars are actually to be considered as fermion stars.
Neutron stars are kind of self-gravitating Fermi gas and the study of this kind of quantum fluid is in the concern of statistical mechanics.
Thus, the neutron star matter, which is usually treated in the classical regime, requires further studies in the quantum regime to be better understood specially under such high pressures and curvatures as those experienced in stars. 

In Physics there are two kinds of quantum statistics: fermion and boson gases. Statistical mechanics is considered by Ingrosso and Ruffini \cite{Ingrosso} in the context of both fermion and boson stars. If the fermion or the boson gas becomes gravitationally bound and stable, the gas will reach a characteristic energy density $\rho$ and pressure $p$ obeying the statistics related to temperature and average energy, such that:

\begin{eqnarray}
p    &=& 16\sqrt{2}S_1 \beta^{5/2} \left[ F_{3/2}(\theta ,\beta) + \frac{\beta}{2}F_{5/2}(\theta ,\beta) \right] 
\label{E1.1}\\
     & &  \nonumber \\
\rho &=& 3\sqrt{2}S_2 \beta^{3/2} \left[ F_{1/2}(\theta ,\beta) + \beta F_{3/2}(\theta ,\beta) \right] \label{E1.2}
\end{eqnarray}
where:
\[\theta =\mu /kT \]
\[\beta = k_B T/mc^2 \]
\[ S_1 =(2s+1)m^4c^5/48\pi^2\hbar^3 \]
\[ S_2 =(2s+1)m^4c^5/6\pi^2\hbar^3 \]
and, $m$ is the mass and $s$ is the spin of the considered particle (fermion or boson). 
The $F$ functions are mathematical tools from statistical mechanics \cite{Pathria}, given by:

\begin{eqnarray}
F_k (\theta ,\beta )  & = & \int_0^{\infty} x^k \left( 1+\frac{\beta x}{2}\right)^{1/2} g(x, \theta )  dx  \label{E1.3} \\ 
     &   &  \nonumber \\ 
g(x, \theta )         & = & \frac{1}{\exp (x-\theta) \pm 1}  \label{E1.4}
\end{eqnarray}
where the sign $+(-)$ refers to fermions (bosons).
The gas presents a chemical potential $\mu$, and $k_B$ is the Boltzmann constant.


\section{Polytropes}\label{s:Polyt}
The study of polytropic stars provided great simplifications concerning the treatment of compact objects.
In thermodynamics polytropes are paths similar to adiabatics, isobarics and isothermals, and families of Lane-Emdem equations can be separated concerning their polytrope indices $n$. This is a very important tool to classify self-gravitating objects and defining their internal energy, gravitational energy and  possible stability \cite{Zeldovich}.

\subsection{Relativistic stars}
In such to consider gas spheres, we use the spherical metric element:
\begin{equation}
ds^{2}=-B(r)d\tau ^{2}+A(r)dr^{2}+r^{2}d\theta ^{2}+r^{2}\sin ^{2}\theta
d\varphi ^{2}  \label{3.1}
\end{equation}

Chandrasekhar proposed that a compact object could be treated as a perfect fluid
described by an energy-momentum tensor:
\begin{equation}
T_{\mu\nu}=p g_{\mu\nu} + \left( p+\rho c^2 \right) u_\mu u_\nu  \label{3.2}
\end{equation}
where $u^\nu$ is the four-velocity of the gas, $p$ is the pressure and $\rho$ is the energy density. Taking units where $c=1$, the four-velocity vectors $u$ obey the relation $u_\mu u_\nu g^{\mu\nu} =-1$, in such that $u_r = u_\theta = u_\phi =0$ and  $u_t = -(g^{tt})^{-1/2} = -\sqrt{B}$. 
The non-vanishing Ricci components, $R_{rr}$,$R_{\theta\theta}$ 
and $R_{tt}$, give rise, respectively, to:

\begin{equation}
\frac{B''}{2B}-\frac{B'}{4B}
\left(  \frac{A'}{A} + \frac{B'}{B} \right) - \frac{A'}{rA}
= - 4\pi G (\rho -p)A   \label{3.3a}  
\end{equation}

\begin{equation}
-1 + \frac{r}{2A}
\left( - \frac{A'}{A} + \frac{B'}{B} \right) + \frac{1}{A}
= - 4\pi G (\rho -p)r^2     \label{3.3b} 
\end{equation}

\begin{equation}
-\frac{B''}{2A} + \frac{B'}{4A}
\left(  \frac{A'}{A} + \frac{B'}{B} \right) - \frac{B'}{rA}
= - 4\pi G (\rho + 3p)B  \label{3.3c}
\end{equation}

It is also useful the equation for the hydrostatic equilibrium:
\begin{equation}
  \frac{B'}{B} = - \frac{2p'}{p+\rho}   \label{3.4}
\end{equation}
and the equation for the total mass inside the star:
\begin{equation}
  {\cal M} = \int_{0}^{R} 4\pi r^2 \rho (r) dr   \label{3.5}
\end{equation}
So, it is possible to define $A(r)$ using the limit to the infinity:
\begin{equation} 
A(r)=\frac{1}{ \displaystyle{
\left(  1-\frac{2G{\cal M}}{r} \right)}  } \label{3.6}
\end{equation}

Equation (\ref{3.3b}) can be rewritten as:
\begin{eqnarray}
-1 + \left[ 1-\frac{2G{\cal M}}{r} \right] 
     \left[ 1-\frac{rp'}{p+\rho} \right] + \frac{2G{\cal M}}{r}
     -4\pi G\rho r^2      \nonumber \\
     = - 4\pi G \left( \rho - p \right) r^2
\label{3.7}
\end{eqnarray}
This last equation can be written as:
\begin{equation}   
-r^2 p'(r) = G{\cal M} \rho    
     \frac{ {\displaystyle \left[ 1+ \frac{p}{\rho} \right]
      \left[ 1+ \frac{4\pi G r^3 p }{\cal M} \right]} }
     {{\displaystyle \left[  1-  \frac{2G{\cal M}}{r} \right]}   }    
\label{3.8}
\end{equation}

This equation can be used to determine pressure evolution in $r$. It can be used in Astrophysics applications within the Newtonian approach after some non-relativistic corrections. Two isentropic special cases are of interest: stars at absolute zero (neutron stars, white dwarf stars and boson stars), and stars in convective equilibrium.

\subsection{Semi-relativistic approach}
When considering the Newtonian case, one can take \cite{Weinberg}:
\begin{equation}   
p \ll \rho
\label{3.9a}
\end{equation}

\begin{equation}   
4\pi r^3 p \ll {\cal M}
\label{3.9b}
\end{equation}

\begin{equation}   
\frac{2G{\cal M}}{r} \ll 1
\label{3.9c}
\end{equation}

Using these approximations in eqn.(\ref{3.8}) simplify to:
\begin{equation}   
-r^2 p' = G {\cal M} \rho 
\label{3.10}
\end{equation}
which by means of eqn.(\ref{3.5}) gives:
\begin{equation}   
\frac{d}{dr}\left[ \frac{r^2}{\rho} p' \right] = 
-4 \pi G r^2 \rho 
\label{3.11}
\end{equation}
This equation could be a linear differential equation (LDE), except 
for the fact that it is to be solved both on $p(r)$ and on $\rho (r)$. 
But, if $p$ and $\rho$ have a relation (known as 'equation of state') then eqn.(\ref{3.11}) becomes
an actual LDE, with solution depending only upon $p$ or $\rho$.
In fact, this equation of state is the polytropic equation:
\begin{equation}   
p= K \rho^\gamma
\label{3.12}
\end{equation}
where $K$ and $\gamma$ are constants. Now one can see that with the adequate boundary conditions, such as $p'(0)=0$ (to keep $\rho (0)$ finite), eqn.(\ref{3.11}) gives $p=p(r)$.

This relation is famous for long time since, for instance, W. Thomson (Lord Kelvin) in 1887 studied the natural stirring produced in a great free fluid mass like the Sun while it is cooling at its surface. 
At Joule's suggestion, Kelvin also studied this natural stirring of a moist atmosphere condensation of vapor in the upward currents of air, which is nowadays of recognized importance in meteorology, known as polytropic change \cite{Chandrasekhar}.  
Following this historical tradition, in Astrophysics any star for which the equation of state takes the form (\ref{3.12}) is called a polytrope \cite{Weinberg}. 
Some typical cases are:  $\gamma =6/5$ is in the range of super large gaseous stars; 
white dwarfs (and generally fermion stars) present $4/3 \leq \gamma  \leq 5/3$, where $\gamma \cong 4/3$ correspond to largest mass white dwarfs and $\gamma \cong 5/3$ to small mass white dwarfs; there are also the incompressible stars  with very high polytropic indexes ($\gamma \rightarrow \infty$).

\subsection{Isothermal gas spheres}

Isothermal gas spheres are important in Astrophysics since it is a starting point to understand composite stars, and also to the study of stars consisting of envelopes with different temperatures \cite{Chavanis}.
For a standard star \cite{Chandrasekhar}, we have from the theorems of the equilibrium of the star:
\begin{equation}
p= \left( \frac{k_B}{\mu H}\right) \rho T + \frac{1}{3} \sigma T^4
\label{3.13}
\end{equation}
where $k_B$ is the Boltzmann constant, $\mu$ is the mean molecular weight, $H$ is the mass of the proton, 
and $\sigma$ is the Stefan-Boltzmann constant.

If the star is in gravitational equilibrium, one can write its equation of state as:
\begin{equation}
p= K\rho + D 
\label{3.14}
\end{equation}
with:
\[ K=\frac{k_B T}{\mu H} \,\,\, , \,\,\,  D=\frac{1}{3} \sigma T^4 \]

Notice that both $K$ and $D$ depends on the temperature. 
Comparing  eqns.(\ref{3.12}) and (\ref{3.14}) we also can learn that the isothermal sphere case
is close to correspond to a polytropic equation with $\gamma =1$, 
if we take $D\rightarrow 0$.


\section{Taylor expanding the equation of state}\label{s:Taylor}

In this article we assume the star is under hydrostatic equilibrium, and using equations 
(\ref{E1.1})-(\ref{E1.2}) one can show that:
\begin{eqnarray}
	p=\left(\frac{16 S_1}{3 S_2}\right) \rho + 8\sqrt{2} S_1 \beta^{7/2}F_{5/2}
	   - 16\sqrt{2} S_1 \beta^{3/2}F_{1/2}
	\label{E4.1}
\end{eqnarray}
which is a polytropic-like equation, unless for the two additional terms at the end. 
It can be compared to equation (\ref{3.12}) if we consider:
\begin{eqnarray}
	p= K\rho^\gamma + C(s,\theta , \beta)
	\label{E4.2}
\end{eqnarray}
where $K$ is constant for a star in equilibrium, and the power factor $\gamma$ 
is sometimes expressed using $\gamma = (1+ 1/n)$ or $n=(\gamma -1)^{-1}$, where $n$ is called the polytropic index.

New physics can be inferred if equations (\ref{3.12}) and (\ref{E4.1}) are generalities of:
\begin{eqnarray}
	p= K\rho^{1+\delta}
	\label{E4.3}
\end{eqnarray}

Some considerations about the results so on.
The function $C$ depends on temperature, but the polytropic equation does not explicitly takes into account
the temperature of the star. Moreover, the case $\gamma =1$ corresponds to an isothermal sphere at constant 
temperature and can be considered as a special situation in Astrophysics, since $n\rightarrow\infty$.
Hence, $\delta$ takes into account deviations from this case (if $\delta =0$ we recover the isothermal sphere).

We can Taylor expand (\ref{E4.3}) around the central density $\rho =\rho _0$:
\begin{eqnarray}
	p=K\left\{ \rho_0^{1+\delta } + (\delta +1)\rho_0^{\delta}\rho - (\delta +1)\rho_0^{1+\delta }
	\right\} + {\cal O} ((\rho -\rho_0 )^2 , \delta ^2 )
	\label{E4.4}
\end{eqnarray}
Thus, up to first order in $\rho -\rho_0$ and $\delta$:
\begin{eqnarray}
	p=\left[ K (\delta +1)\rho_0^{\delta } \right] \rho
	   + K \rho_0^{1+\delta } - K (\delta +1) \rho_0^{1+\delta }
	\label{E4.5}
\end{eqnarray}

A comparison between this last equation and (\ref{E4.1}) brings us to a system of non-linear equations,
which is straightforward solved giving:
\begin{eqnarray}
 \left\{
\begin{array}{lcl}
   \delta &=& 2\alpha -1                             \label{E4.7a} \\
   \rho_0 &=& 3 \sqrt{2}  S_2 \beta^{3/2} F_{1/2}       \label{E4.7b} \\
      K   &=& {\displaystyle \frac{8\sqrt{2}  S_1 \beta^{7/2}F_{5/2}}
                              {\left( 3\sqrt{2}  S_2 \beta^{3/2}F_{1/2}\right)^{2\alpha}} }  \label{E4.7c}
\end{array}
 \right.
\end{eqnarray}
where:
\begin{eqnarray}
 \alpha = \frac{F_{1/2}}{\beta^2 F_{5/2}}  
  \label{E4.6}
\end{eqnarray}

The value of $\delta$ brings relevant information, since one can associate it with the polytropic index, i.e., $\delta = 1/n$. 
Apart of $F_k$ functions one can write: $\alpha \sim m^2 c^4/k_{\rm{B}}^2 T^2$.
Thence, since function $\delta$ is directly related to $\alpha$, which depends on the mass of the particle on consideration, results for bosons and fermions may present large numerical differences.


\section{Results} \label{s:Results}

The aim here is to show that there is a relation between the polytropic model and the statistical mechanics of the gas inside the star. Consequently, in the previous section it has been shown that the parameters $\delta$, $\rho_0$ and $K$, can be related to the temperature and to the chemical potential.
We separate the results for bosons and for fermions in such to compare the behaviour for these two cases. 

To illustrate the parameters behaviour we have used neutrons as fermions, and the $Z^0$ bosons. Their masses, spins and other constants can be recovered, {\em e.g.}, in Particle Data Group \cite{PDG}. Here we consider fermion mass is 938.3MeV/$c^2$ with spin 1/2, and boson mass is $91.19$GeV/$c^2$ with spin 1.

For fermions, Fig.1 shows the plot for the parameter $\delta$ in equation (\ref{E4.7a}) in function of the temperature and the chemical potential. The evaluation of the expression for $\delta$ in this case reveals an inverse dependence in square temperature and a constant of order 10 to 27, {\em i.e.}, $\delta \sim 10^{27} T^{-2}$, which is the dominant term. In this case, we can see that $\delta$ presents large values for the temperature range. The temperature range has been chosen to vary from $T\sim 0$ to $T\sim 3$K, which is the range where significant variations have been observed. The chemical potential has been chosen to vary between $10^{-26}$ and $10^{-25}$ since in this range the integrals present some variation. The statistical mechanics integrals, 
eqs.(\ref{E1.3})-(\ref{E1.4}), are theoretically defined from zero to infinity, but for numerical purposes we used Simpson integration with $x$ variating from $0.1$ to infinity to avoid a  singularity in MAPLE during boson computations.

\begin{figure}
  \includegraphics[angle=270,width=\linewidth ]{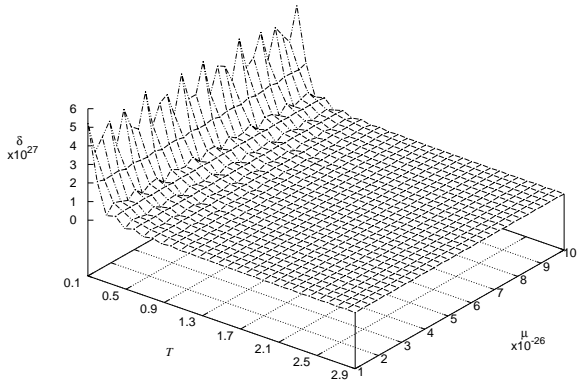} 
\caption{Plot of $\delta$ in function of the temperature $T$ and chemical potential $\mu$ for fermions. Notice that the values of $\delta$ are in order of $10^{27}$, and decreases strongly for higher temperatures. \protect \label{fig1}}
\end{figure}

For bosons, Fig.2 show that $\delta$ there is a similar behaviour. Meanwhile, notice that for low temperatures the function increases to values that are of order $10^{31}$; this is because
the expression for $\delta$ in this case is of order $10^{31} T^{-2}$.
One have to point out that despite fermions and bosons cases present similar displays, the values on $\delta$ axis are very different in order: boson gas can reach higher values, since $Z^0$ mass is larger than the neutron one.

If one uses the so expected Higgs boson ($H^0$, presently under search in CERN's LHC), with expected mass above 114GeV (LEP) \cite{PDG}, the discrepancies are greater. 

\begin{figure}
  \includegraphics[angle=270,width=\linewidth ]{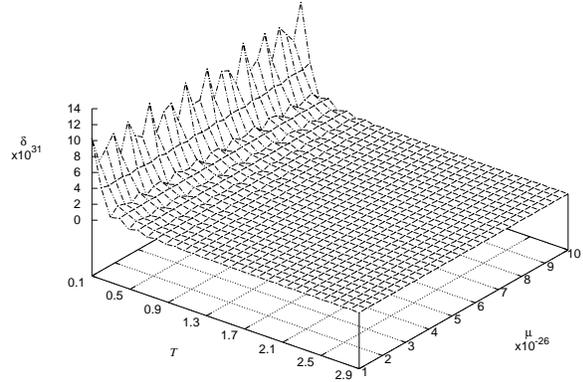} 
\caption{Plot of $\delta$ for bosons. Notice that the values are in 
order $10^{31}$, and decreases as temperature increases. \protect \label{fig2}}
\end{figure}

The behaviour for the parameter $\rho_0$ for fermions is shown in fig.3, and for bosons in fig.4.
Aside the integrals involved, the parameter $\rho_0$ scales like $T^{3/2}$, growing with temperature. (Observe that the two displays present the axis for $T$ and $\mu$ in different direction if compared to fig.1 and fig.2).

\begin{figure}
  \includegraphics[angle=270,width=\linewidth ]{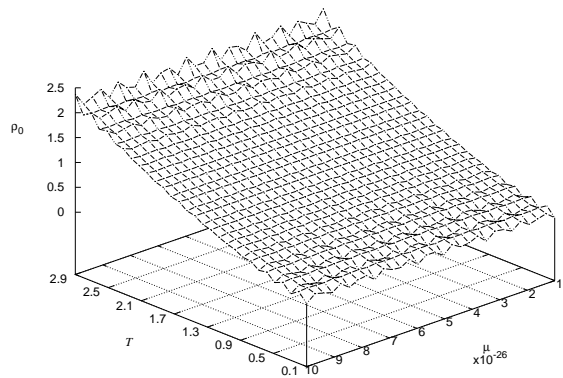} 
\caption{Plot of $\rho_0$ function for fermions. Notice that in this case 
$\rho_0 < 2.5$.  \protect \label{fig3}}
\end{figure}

\begin{figure}
  \includegraphics[angle=270,width=\linewidth ]{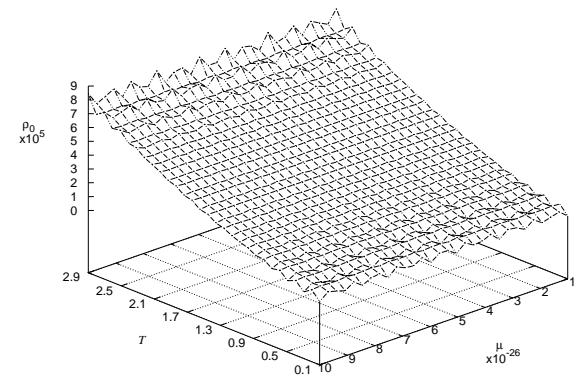} 
\caption{Plot of $\rho_0$ function for bosons. Notice that the values are in 
order $10^{5}$. \protect \label{fig4}}
\end{figure}

In the range of temperature selected for these sample graphics we have obtained nearly vanishing $K$ functions (for both fermions and bosons); meanwhile, for high temperatures we have observed a variating $K$ parameter. 
In the range of temperatures where both $\delta$ and $\rho_0$ have significant values the parameter $K$ is nearly vanishing, and vice versa.
This result for $K$ requires further investigation, since $K \neq 0$ is a necessary condition to avoid a vanishing pressure gas (this is a work in progress).

Anyhow, for any of the numerical ranges tested we observed that $\delta =1/n$ (and $\rho_0$) presented a strong correlation with temperature and chemical potential, within the limits of the expansion used for the equation of state.


\section{Limits and Validity} \label{s:Limits}

The results obtained in the previous section can rise questions concerning the expansion validity for any of the numerical ranges evaluated, since $\delta$ and $\rho_0$ display huge values. In fact,
figures \ref{fig1} to \ref{fig4} form just part of a whole: in those figures only domains where the surface presented pronounced variation are presented; temperature and chemical potential ranges extend to infinity with no limitations up to now. 
Despite having achieved the aim of this work, which is to say showing there is a connection between the polytropic and the isothermal sphere models, we perform a deeper analysis of the polynomial expansion (\ref{E4.5}).

As explained in section~\ref{s:Taylor}, equation (\ref{E4.5}) comes from: 
\begin{eqnarray}
	p &=& K\left\{ \rho_0^{1+\delta } + (\delta +1)\rho_0^{\delta} (\rho -\rho_0 ) 
	          + \frac{\delta (\delta -1)}{2} \rho_0^{\delta -1}(\rho -\rho_0 )^2   \right. \nonumber \\
	  & & \left.   + \frac{\delta (\delta^2 -1)}{6} \rho_0^{\delta -2}(\rho -\rho_0 )^3 + \cdots  \right\}
	\label{E6.1}
\end{eqnarray}
which is a Taylor expansion of (\ref{E4.3}) around the central density $\rho =\rho _0$. 
Two points are important to keep the validity for this expansion: terms $(\rho -\rho_0 )^2 \rightarrow 0$ and $\delta \rightarrow 0$. 

The first condition is a necessary one otherwise parts of the third term in eq.(\ref{E6.1}) mix up with those from the previous term. Moreover, the first condition can be interpreted as $\rho (r) \simeq\rho_0$, which is a reasonable demand when considering an isothermal gas sphere during the formation of a star, for instance.
The second condition is necessary since there are terms in $\delta^2$ that must be neglected in such eq.(\ref{E4.5}) can be considered an useful approximation and to guarantee its resemblance to eq.(\ref{E4.1}).

In figures \ref{fig1} to \ref{fig4} we can see that $\delta$ displays huge values and it is questionable if the comparison that results in the present model is valid. But, as remarked above the values of $T$ and $\mu$ extend to infinity. Thence, we can look for regions in the domain $T \times\mu$ where $\delta \rightarrow 0$. 

Since $\delta = 2\alpha -1$, the limit $\delta \rightarrow 0$ yields $\alpha\rightarrow 1/2$, or:
\[
\left( \alpha -\frac{1}{2} \right) \rightarrow 0 .
\]

Numerically we defined a region where this limit is achieved with a certain precision $\epsilon$:
\begin{eqnarray}
 	\left| \alpha -\frac{1}{2} \right| < \epsilon 
		\label{E6.2}
\end{eqnarray}
and we have used this condition to search only for points that are within this precision.
If we can find such points we perform an existence condition. 
Indeed we were able to find such points for both configurations, boson and fermion stars.
This would be expected since $\delta$ decreases as $T$ and $\mu$ increases. 

In fact, considering fermions to be neutrons, then $\alpha$ in eq.(\ref{E4.6}) is given by:

\begin{eqnarray}
  \alpha_{\rm Fermi}=
 	0.1186\times 10^{27} 
 	\frac{\int_0^{\infty} \frac{ \sqrt{x}\sqrt{1+0.4592\times 10^{-13}Tx} }
 	                           {\exp\left( x- \frac{0.7243\times 10^{23}\mu}{T}  \right) +1 }   }
 	     {T^2 \int_0^{\infty} \frac{ x^{5/2}\sqrt{1+0.4592\times 10^{-13}Tx} }
 	                           {\exp\left( x- \frac{0.7243\times 10^{23}\mu}{T}  \right) +1 } }
		\label{E6.3}
\end{eqnarray}
which shows that, to obtain the required limit we need 
\[ 
\alpha\sim \frac{10^{27}}{T^2} \rightarrow \frac{1}{2}, 
\]
and that yields temperatures such as $T\sim 10^{13}$K. Since eq.(\ref{E6.3}) demand a big computational effort we use this analytical view to search for points in the vicinity of this temperature (for fermion configurations the program spent about 5 hours and for bosons about 18 hours, using standard computers).
In the case of $Z^0$ bosons we have obtained:
\begin{eqnarray}
  \alpha_{\rm Bose}=
 	0.1120\times 10^{31} 
 	\frac{\int_0^{\infty} \frac{ \sqrt{x}\sqrt{1+0.4725\times 10^{-15}Tx} }
 	                           {\exp\left( x- \frac{0.7243\times 10^{23}\mu}{T}  \right) -1 }   }
 	     {T^2 \int_0^{\infty} \frac{ x^{5/2}\sqrt{1+0.4725\times 10^{-15}Tx} }
 	                           {\exp\left( x- \frac{0.7243\times 10^{23}\mu}{T}  \right) -1 } }
		\label{E6.3}
\end{eqnarray}
and for that the required limit for $\alpha$ is achieved with temperatures $T\sim 10^{13}$K.

\begin{figure}
  \includegraphics[angle=270,width=\linewidth ]{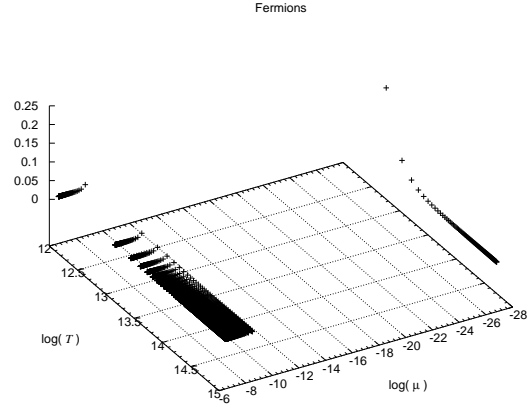} 
\caption{Search for points attending eq.(\ref{E6.2}) with $\epsilon =0.5$. Crosses correspond to points where $\delta \rightarrow 0$ for the fermions case. Logarithm is base 10.
 \protect \label{fig5}}
\end{figure}

Fig.~\ref{fig5}, then, shows the points obtained for the fermions case. Crosses then represent points attending eq.(\ref{E6.2}) with $\epsilon =0.5$, $\mu$ variating from $1\times 10^{-27}$ to $2\times 10^{-7}$ with step $2\times 10^{-9}$, and temperature variating from $1\times 10^{12}$ to $9\times 10^{14}$ with step $9\times 10^{12}$, giving ten thousand points to be tested. Hopefully, 
we obtained some points with $\epsilon \leq 0.5$ that are those for which the model can
serve to the desired purpose since $\delta$ is small.
Notice that we use a log-log plot due to the great discrepancy for the values.

\begin{figure}
  \includegraphics[angle=270,width=\linewidth ]{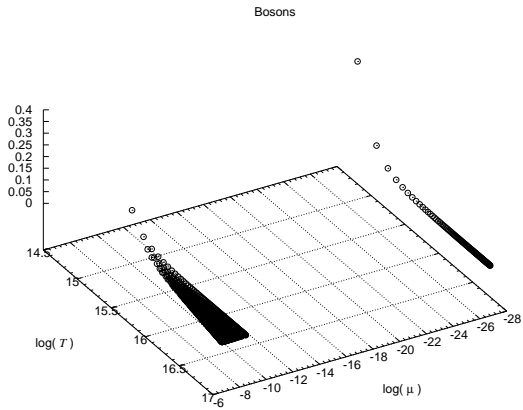} 
\caption{Search for points attending eq.(\ref{E6.2}) with $\epsilon =0.5$. Circles correspond to points where $\delta \rightarrow 0$ for the bosons case. Logarithm is base 10.
 \protect \label{fig6}}
\end{figure}

Fig.~\ref{fig6} shows the points obtained for the bosons case. Circles represent points attending eq.(\ref{E6.2}) with $\epsilon =0.5$, $\mu$ variating from $1\times 10^{-27}$ to $2\times 10^{-7}$ with step $2\times 10^{-9}$, but temperature variating from $1\times 10^{14}$ to $9\times 10^{16}$ with step $9\times 10^{14}$, giving again ten thousand points to be tested. For bosons we also observed the existence of points for which the model apply within the precision $\epsilon$.

\begin{figure}
  \includegraphics[angle=270,width=\linewidth ]{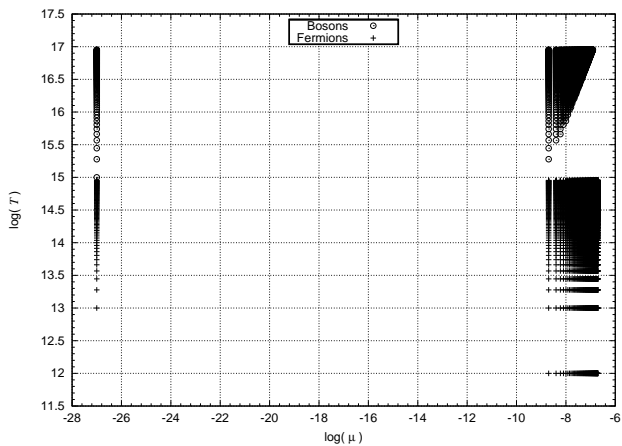} 
\caption{Merged plot of both cases, while searching for points attending eq.(\ref{E6.2}) with $\epsilon =0.5$. Spits at the top correspond to bosons and spits at the bottom correspond to fermions.
 \protect \label{fig7}}
\end{figure}

Fig.~\ref{fig7} shows both cases as seen from above. One can see some clouds of points (that due to the softness of the function grouped like `spits'). The number of points obtained for fermionic configurations was greater than the bosonic ones, in the limits tested. From this plot one can see that there is a difference at the characteristic temperatures for bosons and fermions, and we can define different 'regimes' where the model approaches correctly polytropes to isothermal gas spheres.

Once one can rely on the existence of regions where $\delta\rightarrow 0$ there remains the following question: are corresponding values of central density $\rho_0$ comparable to those obtained in the literature?
A rapid test can be made by taking two sample points in the clouds of the figure \ref{fig7} for each case. We chose $\log(\mu )=-8$, and temperatures $\log ( T )\cong 14.5$ for fermions and $\log ( T )\cong 16.5$ for bosons.

\begin{table*}
 \centering
 \begin{minipage}{140mm}
  \caption{Central densities $\rho_0$: sample cases for boson and fermion numerically obtained by the present model compared to those cited in the literature (section~\ref{s:Limits}). Quantities dimensions are: $\mu$ in Joules, $T$ in Kelvins, and $\rho_0$ in ${\rm kg}/{\rm m}^{3}$ (fifth row).  Attention: there are differences between those $\rho_0$ units. Refer to the text for further information.  \label{tab1}}
  \begin{tabular}{l|ccccl}
  \hline \\
Case &      $\mu$        &           $T$        & $\rho_0$             &        $\rho_0$    \\
     &                   &                   &                         &    (Correct Dimensions)   \\ 
 \hline \\
 Fermions & $1\times 10^{-8}$ & $3.16\times 10^{14}$ &  $6.5195\times 10^{22}$ & 
           5.87 $\times 10^{39}$                 \\
\hline \\
 Bosons   & $1\times 10^{-8}$ & $3.16\times 10^{16}$ &  $3.7262\times 10^{30}$ & 
          3.35 $\times 10^{47}$                  \\
 \hline 
\end{tabular}
\end{minipage}
\end{table*}

In Table~\ref{tab1} numerical values computed for mass density are directly listed in the fourth row, 
for the described sample cases. But, during the treatment of relativistic stars, since eq.(\ref{3.2}), numerical values for $\rho$ (and $\rho_0$) are in lack of a $c^2$ factor. 
Observe that this gives the correct units for mass density: notice that $\rho$ has the same units as $S_2$,
and once $S_2$ is expressed in MKS-units
${\rm kg}\, {\rm m}^{-1}\, {\rm s}^{-2}$ the adequate correction in units of $c^2$ yields 
${\rm kg}\, {\rm m}^{-3}$ as expected.
Central densities on the fourth row are computed directly  using eq.(\ref{E4.7b}) and are not in correct dimensions.

From Table~\ref{tab1} one can see that the sample values so obtained (last row) give rise to densities that are in accordance to those expected in reference texts. 
In fermion stars densities central densities are  of order $10^{40}$kg/m$^3$ \cite{Zeldovich, Ruffini}. 
In boson stars central densities are given by $\rho_0 \sim m^2/4\pi G$ and are of order $10^{48}$kg/m$^3$ \cite{Gleiser2, Gleiser}.


\section{Discussion}

In this paper we show that there is a relation between the statistical mechanics of a self-gravitating gas sphere model and the polytropic model; for short, we propose the terminology polytropic isothermal gas spheres.
We start with the statistical mechanics treatment given for fermion and boson gases \cite{Ingrosso} and find an equivalent equation of state that is compared with the polytropic model. We observe that this relation is similar to isothermal gas spheres, i.e., a polytrope with $\gamma =1$. The main result starts in section \ref{s:Taylor}, where a polynomial expansion is performed and in first order in $\rho$ one can find the suggested correspondence for the polytropic model parameters: $K$, $\rho_0$ and $\delta = 1/n$.
To illustrate our results we use neutrons as fermions and $Z^0$ bosons. The graphics for $\delta$ for fermions and bosons are similar in shape, but very different in values range, since $\alpha$ values show up remarkable differences due to the masses of the particles.


Limits of this approach are the following. 
The main result on section \ref{s:Taylor} is obtained only up to first order. But, we can see this matches the models in use for isothermal gas spheres. Thence, it is not expected remarkable differences if one try a larger polynomial expansion.

We consider only non-charged particles to avoid further interactions. Probably, new physics can be extracted considering charges particles like protons, and $W^{\pm}$ bosons or $H^{\pm}$ Higgs bosons.  

One advantage in this picture is that eqs.(\ref{E4.7b}) are describing the parameters dependence in temperature and chemical potential, and there is no need to separate the cases for low temperature and high temperature \cite{Ingrosso}. Initially we focused at low temperature range, closer to the (environment) cosmic background temperature, since the considered compact objects like neutron stars and boson stars are theoretically shaped using zero temperature. 

In order to preserve  model validity we performed an analysis searching for regions (temperature versus chemical potential) where the expansion is adequate. We found those regions and corresponding central densities are similar to those known for fermion and boson stars.
Future work could use the set of equations similar to (\ref{E4.7b}), with a higher order approximation in the expansion, to detail the behaviour of the parameters for larger ranges of temperature and chemical potential which could be useful for a more expanded set of stellar objects and structures.

Merafina \cite{Merafina} points out that semi-degenerate configurations have had a quite satisfactory treatment inside the bulge of the star, but for the outer shell, where there is no border, the thermodynamical quantities expansions fail. This problem can possibly be circumvented if one uses the polytropic approach to describe the fermionic stars (neutrons stars and white dwarf stars) and boson stars.


The overall conclusion is that there is a strong correspondence between polytropic and gas sphere thermal models, with an evident correlation within their parameters, e.g., the polytropic index $n$, and possible temperatures, composition and chemical potentials. 
This is exactly what is expected from the theory of stellar interiors \cite{Chandrasekhar, Zeldovich}, but has never been reported before. Hence, in this paper we present an alternative to link parameters, and for studying their significant ranges in function of temperature and chemical potential. In special, the analysis of $n\rightarrow\infty$ \cite{Natarajan}, and the envisage of some possible implications for cloud condensation \cite{Honda} is of potential interest.

\section*{Acknowledgments}

We wish to thank L. Ph. Vasconcelos for proof-reading.
We would like to thank  A.L.A. Fonseca, M. D. Maia and E. A. Asano for
their encouragement and suggestions. Special thanks to the referee since his/her 
clever questions led to a more detailed analysis of expansion limits of the model.
This work was a complementary research activity of NAS -- 
Nucleo de Astronomia de Santarem (Astronomy Group of Santarem).


\end{document}